\documentclass[12pt]{article}

\usepackage{graphicx}
\usepackage{amsmath}
\usepackage{xcolor}
\usepackage{amssymb}
\usepackage{hyperref}
\usepackage{authblk}

\usepackage[a4paper, margin=1in]{geometry}

\title{Detection of Dark Matter using levitated nanoparticles within a Bessel-Gaussian beam via Yukawa coupling}

\author[1]{Iftekher S. Chowdhury}
\author[2]{Binay Prakash Akhouri}
\author[5]{Shah Haque}
\author[1,3]{Martin H. Bacci}
\author[1,4,5]{Eric Howard}

\affil[1]{Department of Physics and Astronomy, Macquarie University, Sydney, NSW, 2109, Australia}
\affil[2]{Department of Physics, Suraj Singh Memorial College, Ranchi University, Ranchi, Jharkhand, India}
\affil[3]{Maritime and Polytechnic University College, Frederikshavn, Denmark}
\affil[4]{Swinburne University, Sydney, Australia} 
\affil[5]{Southern Cross Institute, School of Computer Science, Sydney, Australia} 

\date{\today}

\begin{document}

\maketitle

\begin{abstract}
\begin{abstract}
We present a novel experimental approach to detect dark matter by probing Yukawa interactions, commonly referred to as a fifth force, between dark matter and baryonic matter. Our method involves optically levitating nanoparticles within a Bessel-Gaussian beam to detect minute forces exerted by potential dark matter interaction with test masses. The non-diffracting properties of Bessel-Gaussian beams, combined with feedback cooling techniques, provide exceptional sensitivity to small perturbations in the motion of the nanoparticles. This setup allows for precise control over trapping conditions and enhances the detection sensitivity to forces on the order of \(10^{-18}\) N. We explore the parameter space of the Yukawa interaction, focusing on the coupling strength (\(\alpha\)) and interaction range (\(\lambda\)), and discuss the potential of this experiment to place new constraints on dark matter couplings, complementing existing direct detection methods.
\end{abstract}

\end{abstract}

\section{Introduction}

Understanding the nature of dark matter remains one of the greatest challenges in modern physics. Although dark matter is thought to constitute approximately 27\% of the universe’s mass-energy content, its direct detection has proven elusive. Unlike baryonic matter, dark matter does not interact with electromagnetic radiation, making it invisible to telescopes and other traditional detection methods. Instead, its existence has been inferred through gravitational effects on visible matter, radiation, and the large-scale structure of the universe.

The most convincing evidence for dark matter comes from several key astrophysical observations. For example, the rotation curves of galaxies suggest the presence of a substantial amount of unseen mass, as stars in galaxies rotate at higher speeds than can be explained by the visible matter alone. Similarly, measurements of gravitational lensing, particularly in galaxy clusters, show that the mass necessary to bend light far exceeds the amount of detectable luminous matter. Observations of the cosmic microwave background, as well as large-scale surveys of galaxy distributions, further support the hypothesis that a significant component of the universe is composed of dark matter. Despite this overwhelming evidence, the specific nature of dark matter remains unknown, and identifying the particle or particles responsible is one of the key goals of modern physics.

Traditionally, two leading candidates have dominated the search for dark matter: weakly interacting massive particles (WIMPs) and axions. WIMPs, predicted by several extensions of the Standard Model of particle physics, would interact with ordinary matter through weak nuclear forces and are considered to be ideal candidates for direct detection experiments. Axions, on the other hand, are low-mass particles that emerge as a solution to the strong CP problem in quantum chromodynamics. Both WIMPs and axions are expected to interact weakly with baryonic matter, making them detectable through either direct search experiments or indirect observations such as gamma rays or cosmic rays. Despite extensive efforts, including underground detectors, collider searches, and astrophysical measurements, there has been no definitive evidence for WIMPs or axions, prompting the exploration of alternative models for dark matter.

One such alternative approach involves the consideration of force candidates between dark matter and baryonic matter. These models propose that dark matter could interact with ordinary matter via a force similar to gravity, but with a different range and strength. A promising candidate for this interaction is the Yukawa potential, sometimes referred to as a "fifth force." This potential modifies the standard gravitational force by adding an exponentially decaying term, characterized by two parameters: the coupling strength \(\alpha\), which determines the strength of the interaction relative to gravity, and the interaction range \(\lambda\), which defines the characteristic length scale over which the interaction remains significant. Unlike short-range interactions, such as those expected from WIMPs or axions, a Yukawa-type interaction could manifest over macroscopic distances, providing a new avenue for the experimental detection of dark matter.

The Yukawa interaction is particularly intriguing because it introduces the possibility of dark matter influencing baryonic matter over large distances. In astrophysical systems, this interaction could lead to observable deviations from Newtonian dynamics, for instance, in the outer regions of galactic halos or in galaxy clusters. These deviations could provide an alternative explanation for certain phenomena currently attributed to dark matter, such as the flattening of galaxy rotation curves. On the laboratory scale, the Yukawa interaction may manifest as a subtle perturbation in the motion of a test mass, such as a levitated nanoparticle. The detection of such weak forces requires highly sensitive experimental setups, capable of isolating the test mass from environmental noise and tracking its motion with high precision.

Recent developments in optomechanical systems, particularly the use of optically levitated nanoparticles, offer a unique platform for probing weak interactions such as the Yukawa force. By trapping nanoparticles in vacuum using laser beams, it is possible to achieve an extremely sensitive detection system that is isolated from external disturbances such as air drag and thermal noise. The ability to track the motion of these nanoparticles with nanometer precision makes them ideal candidates for detecting the minute forces expected from a Yukawa interaction. In addition, the use of feedback cooling techniques allows the system to reduce the thermal motion of the particle, further increasing the sensitivity of the experiment by bringing the nanoparticle close to its motional ground state.

In this work, we employ a Bessel-Gaussian beam to trap nanoparticles in a highly controlled environment. Bessel-Gaussian beams have several advantages over traditional Gaussian beams, most notably their non-diffracting nature, which allows them to maintain a stable intensity profile over long distances. This is crucial for maintaining a consistent trapping force on the nanoparticle, which is necessary for detecting weak external forces such as those arising from the Yukawa interaction. Furthermore, the combination of Bessel-Gaussian trapping with feedback cooling techniques provides the level of precision required to detect forces on the order of \(10^{-18}\) N, making this setup highly suited for exploring dark matter interactions.

The goal of this experiment is to explore the feasibility of detecting dark matter via Yukawa interactions by measuring the perturbations in the motion of an optically levitated nanoparticle. By analyzing these perturbations, we aim to place constraints on the coupling strength \(\alpha\) and interaction range \(\lambda\) of the Yukawa interaction. This work offers a complementary approach to traditional dark matter detection methods, such as those searching for WIMPs and axions, by focusing on interactions that could be probed using precision optomechanical techniques. The combination of advanced trapping technology, noise mitigation strategies, and precision measurement provides a new avenue for investigating the fundamental nature of dark matter. 

\section{Theoretical framework}

The Yukawa interaction, commonly referred to as the fifth force, provides an extension to the traditional gravitational interaction by introducing a new force that decays exponentially with distance. This interaction is motivated by the hypothesis that dark matter may interact with baryonic matter through more than just gravity, potentially via a scalar or vector field. The Yukawa potential introduces two essential parameters: the coupling constant \(\alpha\), which measures the relative strength of this interaction compared to gravity, and the interaction range \(\lambda\), which defines the characteristic distance over which the force remains significant.

In Newtonian mechanics, the gravitational potential energy \(\Phi_{\text{grav}}(r)\) for a point mass \(M\) at a distance \(r\) from a test mass \(m\) is given by the well-known expression:
\[
\Phi_{\text{grav}}(r) = - \frac{GMm}{r},
\]
where \(G\) is the universal gravitational constant. This potential gives rise to the inverse-square law for gravitational forces, governing interactions over macroscopic scales, from planetary motion to galaxy dynamics. However, the Yukawa interaction modifies this potential by adding a new term that decays exponentially with distance. The modified potential takes the form:
\[
\Phi_{\text{fifth}}(r) = -\alpha \frac{GMm}{r} e^{-r/\lambda},
\]
where \(\alpha\) is the dimensionless coupling constant and \(\lambda\) is the characteristic range of the Yukawa interaction. The exponential decay term \(e^{-r/\lambda}\) introduces a characteristic length scale \(\lambda\), beyond which the Yukawa force rapidly diminishes. 

For distances \(r \lesssim \lambda\), the Yukawa interaction can significantly modify gravitational behavior, while for distances \(r \gg \lambda\), the Yukawa interaction becomes negligible. The total potential acting between two masses is now given by the combination of the gravitational and Yukawa contributions:
\[
\Phi_{\text{total}}(r) = - \frac{GMm}{r} \left( 1 + \alpha e^{-r/\lambda} \right).
\]
Thus, the Yukawa potential introduces a deviation from the standard inverse-square law of gravitation at distances smaller than \(\lambda\).

The force associated with the Yukawa-modified potential can be derived by taking the gradient of the total potential \(\Phi_{\text{fifth}}(r)\). For the gravitational potential alone, the force is:
\[
\mathbf{F}_{\text{grav}}(r) = - \nabla \Phi_{\text{grav}}(r) = \frac{GMm}{r^2} \hat{r},
\]
where \(\hat{r}\) is the radial unit vector. In the case of the Yukawa-modified potential, the total force becomes:
\[
\mathbf{F}_{\text{fifth}}(r) = - \nabla \Phi_{\text{fifth}}(r) = \alpha \frac{GMm}{r^2} e^{-r/\lambda} \left( 1 + \frac{r}{\lambda} \right) \hat{r}.
\]
This expression reveals two key regimes of interest:
\begin{itemize}
    \item \textbf{Short-range regime (\(r \ll \lambda\)):} In this regime, the exponential term \(e^{-r/\lambda} \approx 1\), and the Yukawa force behaves similarly to the gravitational force but with an additional scaling factor \(\alpha\):
    \[
    \mathbf{F}_{\text{fifth}}(r) \approx \alpha \frac{GMm}{r^2}.
    \]
    \item \textbf{Long-range regime (\(r \gg \lambda\)):} In this regime, the exponential term \(e^{-r/\lambda}\) rapidly decays, and the Yukawa force is exponentially suppressed:
    \[
    \mathbf{F}_{\text{fifth}}(r) \approx \alpha \frac{GMm}{r^2} e^{-r/\lambda}.
    \]
\end{itemize}
These regimes emphasize the critical role of the range parameter \(\lambda\), which determines where the Yukawa force becomes negligible and where it significantly modifies the gravitational interaction.

The gravitational potential energy between two masses \(M\) and \(m\) separated by a distance \(r\) is traditionally expressed as:
\[
U_{\text{grav}} = - \frac{GMm}{r}.
\]
With the inclusion of the Yukawa potential, the total potential energy becomes:
\[
U_{\text{total}} = - \frac{GMm}{r} \left( 1 + \alpha e^{-r/\lambda} \right).
\]
The extra term \(e^{-r/\lambda}\) modifies the gravitational potential by introducing a range-dependent exponential suppression, which causes deviations from the inverse-square law at short distances.

To further understand the interaction's behavior, consider the force exerted between the two masses. The classical gravitational force is given by:
\[
\mathbf{F}_{\text{grav}} = - \nabla U_{\text{grav}} = \frac{GMm}{r^2}.
\]
With the Yukawa modification, the total force becomes:
\[
\mathbf{F}_{\text{total}} = - \nabla U_{\text{total}} = \frac{GMm}{r^2} \left( 1 + \alpha e^{-r/\lambda} \left( 1 + \frac{r}{\lambda} \right) \right).
\]
In the short-distance regime (\(r \ll \lambda\)), the force simplifies to:
\[
\mathbf{F}_{\text{total}}(r \ll \lambda) \approx \alpha \frac{GMm}{r^2},
\]
suggesting a rescaling of the gravitational force by a factor of \(1 + \alpha\). In the long-range regime (\(r \gg \lambda\)), the exponential suppression term dominates, and the force reduces to:
\[
\mathbf{F}_{\text{total}}(r \gg \lambda) \approx \frac{GMm}{r^2}.
\]
This shows that for distances beyond \(\lambda\), the Yukawa contribution becomes negligible, and the force approaches the standard gravitational form.

The Yukawa interaction provides a potential mechanism for explaining certain observed discrepancies between predictions made by the cold dark matter (CDM) model and astrophysical observations, particularly in the outer regions of galactic halos or galaxy clusters where dark matter dominates the mass distribution. In the standard CDM model, gravitational forces alone may not fully account for the observed velocity dispersions of stars and galaxies. The additional Yukawa force could help bridge this gap by enhancing the effective gravitational force at short distances.

For example, in galaxy clusters, the interaction could modify the gravitational potential on the outskirts, leading to small deviations in the observed velocity dispersion of galaxies. These effects, while subtle, could be detectable with precise measurements of galactic dynamics and provide new insights into the nature of dark matter and its interaction with baryonic matter.

However, due to the weak nature of the Yukawa interaction and the rapid suppression of the force at large distances, its effects are challenging to detect in astrophysical observations. For distances \(r \gg \lambda\), the Yukawa interaction has little to no effect, making it difficult to observe deviations from the standard gravitational behavior at galactic or larger scales. This motivates the design of highly sensitive laboratory experiments to detect signatures of the Yukawa force at smaller distances, such as sub-millimeter scales.

The Yukawa interaction introduces an important theoretical framework for studying deviations from the inverse-square law, and the resulting modifications to gravitational potential and forces are of great interest in both fundamental physics and cosmology. Expanding on the mathematical structure, the interaction can be modeled in various theoretical frameworks, including scalar-tensor theories and extensions of general relativity. These models often introduce additional scalar or vector fields that mediate the Yukawa interaction, modifying the gravitational field equations accordingly.

The modified Poisson equation for gravity in the presence of a Yukawa interaction is given by:
\[
\nabla^2 \Phi_{\text{total}}(r) = 4 \pi G \rho \left( 1 + \alpha e^{-r/\lambda} \right),
\]
where \(\rho\) is the mass density. The exponential term reflects the contribution of the Yukawa potential to the total gravitational potential.

In more advanced formulations, the coupling constant \(\alpha\) may itself vary with distance or depend on additional parameters, introducing further complexity to the model. Likewise, the range parameter \(\lambda\) may be constrained by experimental data, providing limits on the scale at which the Yukawa interaction becomes significant. These theoretical considerations offer a fertile ground for future research, both in high-energy physics and precision cosmology.

The Yukawa interaction provides a compelling modification to the standard gravitational theory, introducing an additional short-range force that decays exponentially with distance. This interaction could offer insights into the nature of dark matter and its interactions with baryonic matter, as well as provide explanations for discrepancies observed in galactic dynamics. The mathematical framework, characterized by the coupling constant \(\alpha\) and the interaction range \(\lambda\), governs the behavior of this force across different distance scales. While the Yukawa interaction remains challenging to detect, especially at large scales due to its exponential suppression, it presents exciting possibilities for laboratory experiments designed to probe short-range forces. The theoretical models underpinning this interaction continue to offer new avenues for understanding the fundamental forces that govern the universe.

\section{Optical Trapping with Bessel Beams}

Optical trapping relies on the interaction between the electric field of a light beam and a dielectric particle. In traditional Gaussian beams, the diffraction limits the stability of the trap over long distances. However, Bessel-Gaussian beams provide a non-diffracting solution, allowing for stable confinement over extended propagation distances. A Bessel beam is characterized by its ability to maintain its intensity profile, owing to the superposition of plane waves traveling along a conical surface. This feature makes it ideal for levitation experiments.

The electric field of an ideal Bessel beam in cylindrical coordinates \((r, \phi, z)\) can be expressed as:
\[
E(r, z) = E_0 J_0(k_r r) e^{i k_z z},
\]
where \(J_0\) is the zeroth-order Bessel function of the first kind, \(k_r\) is the radial component of the wave vector, and \(k_z\) is the axial component. The total wave vector is related by \(k = \sqrt{k_r^2 + k_z^2}\).

In optical trapping, the gradient force is responsible for drawing the particle towards the region of highest intensity, typically the center of the beam. The gradient force arises from the interaction of the electric field gradient with the induced dipole moment of the particle. In the Rayleigh regime, where the particle size is much smaller than the wavelength of the trapping light, the gradient force \(F_{\text{grad}}\) can be written as:
\[
F_{\text{grad}} = \frac{1}{2} \alpha \nabla |E|^2,
\]
where \(\alpha\) is the polarizability of the particle, and \(E\) is the electric field of the optical beam. For a Bessel beam, the intensity varies radially, leading to a strong restoring force towards the center.

The gradient force is most significant near the high-intensity core of the Bessel beam, where the field gradient is steepest. The optical trap stiffness \( \kappa \), which characterizes the strength of the trap, depends on the power of the trapping laser and the beam waist. For a particle trapped at the beam center, the radial stiffness can be expressed as:
\[
\kappa_r = \frac{2 \pi n P}{c w_0^2},
\]
where \(P\) is the laser power, \(n\) is the refractive index of the surrounding medium, and \(w_0\) is the waist of the beam.

The scattering force originates from the momentum transfer between the incident photons and the particle. Unlike the gradient force, which acts towards the beam center, the scattering force pushes the particle in the direction of beam propagation. For a Bessel beam, the scattering force is given by:
\[
F_{\text{scatt}} = \frac{n P \sigma}{c},
\]
where \(\sigma\) is the scattering cross-section of the particle and \(c\) is the speed of light. The scattering force acts along the \(z\)-axis, tending to push the particle out of the trap.

In the context of dark matter detection, a Yukawa-type interaction introduces a long-range force that could modify the trapping dynamics of the nanoparticle. 
This force could cause detectable perturbations in the motion of the nanoparticle within the optical trap, manifesting as small periodic displacements, which can be measured with high precision.

For stable trapping, the total force acting on the particle must be zero at equilibrium. This condition provides:
\[
F_{\text{grad}} + F_{\text{scatt}} + F_{\text{Yukawa}} = 0.
\]
By balancing the laser power and the optical trap geometry, it is possible to achieve a configuration where small external forces, such as those from dark matter, can induce detectable displacements.

The trapping potential experienced by the nanoparticle is typically harmonic for small displacements from the equilibrium position. The total force acting on the nanoparticle is then the sum of the optical trapping force and any external forces, such as those arising from the Yukawa interaction. 
The equation of motion for a levitated nanoparticle subject to both optical forces and the Yukawa interaction can be written as:
\[
m \ddot{x} + \gamma \dot{x} + kx = F_{\text{Yukawa}}(r),
\]
where:
\begin{itemize}
    \item \(m\) is the mass of the nanoparticle,
    \item \(\gamma\) is the damping coefficient accounting for optical and thermal damping,
    \item \(k\) is the stiffness of the optical trap, and
    \item \(F_{\text{Yukawa}}(r)\) is the external force due to the Yukawa interaction.
\end{itemize}
In this equation, \(x\) represents the displacement of the nanoparticle from its equilibrium position, and \(\gamma \dot{x}\) captures the effects of damping due to thermal noise and other dissipative processes.

Equilibrium between the scattering, gradient, and gravitational forces ensures that the particle remains trapped in its equilibrium position. External forces, such as Yukawa forces, can cause the particle to shift from this equilibrium. These forces modify the system’s dynamics, potentially leading to detectable shifts in the particle's position. The sensitivity of the system can be further increased by reducing the optical trap stiffness, which weakens the trapping potential and amplifies the sensor's response to external forces.

The intensity of the initial Gaussian beam is gradually reduced until its influence on the particle is negligible. The Gaussian beam is then repurposed for tracking the particle’s motion. In this configuration, the scattering forces cancel out, and the dipole force along the z-axis enables the system to be highly sensitive to external perturbations, including potential Yukawa forces.

Bessel beams provide a unique and advantageous configuration for optical trapping, particularly in experiments aimed at detecting weak forces such as those associated with dark matter. By carefully balancing the gradient and scattering forces, it is possible to achieve a highly sensitive trap capable of detecting small perturbations in the motion of levitated nanoparticles. The introduction of Yukawa-type interactions offers an exciting possibility for dark matter detection, with the potential to probe new parameter spaces in dark matter research.

\section{Experimental setup and Methodology}

In laboratory settings, experiments involving optically levitated nanoparticles offer an excellent platform for detecting the Yukawa interaction. These nanoparticles can be isolated in a vacuum using laser trapping techniques, where their motion is monitored with high precision. The optical trapping force that confines the nanoparticle is generated by the interaction between the nanoparticle and the laser field, and any deviation from the expected force could signal the presence of a new Yukawa interaction.

The proposed experiment is aimed at detecting the effects of Yukawa forces on an optically levitated particle, or probe, trapped in a Gaussian-Bessel beam. The probe will be levitated in a high-vacuum environment to minimize external noise and maximize sensitivity. To further enhance the system's precision, we propose using a single-laser feedback cooling mechanism. This setup utilizes one laser both to probe the particle and to regulate its motion. The feedback system continuously monitors the probe’s oscillatory movement and applies corrective forces to dampen its motion, thereby reducing thermal fluctuations.

The experimental process involves three key stages. First, levitation is achieved through the interaction of scattering force, gradient force, and gravity. The particle is initially trapped using a highly focused Gaussian beam. This beam propagates along the x-axis, providing an upward scattering force that counteracts the downward pull of gravity. The gradient force, generated by the focused beam, must be strong enough to overcome both gravitational and thermal motion, ensuring stable confinement at the beam’s focal point in three dimensions.

In a controlled experimental setting, such as one involving optically levitated nanoparticles, the sensitivity to weak forces like the Yukawa interaction can be significantly enhanced. Optically levitated nanoparticles provide an ideal platform for detecting such forces, as they can be trapped in a vacuum using focused laser beams, isolated from most environmental noise sources, and monitored with high precision. The optical trapping force that confines the nanoparticle is generated by the interaction between the nanoparticle and the laser field. The optical forces acting on the nanoparticle include the scattering force, arising from photon momentum transfer, and the gradient force, which depends on the spatial variation of the laser intensity.

The nanoparticle's displacement due to the Yukawa interaction is expected to be small but measurable. The key to detecting these small displacements lies in the system's sensitivity, which can be enhanced through noise reduction techniques such as feedback cooling. Feedback cooling works by continuously measuring the position of the nanoparticle and applying a compensating force that opposes its motion, effectively damping its thermal fluctuations and reducing its effective temperature. By minimizing the thermal motion of the nanoparticle, feedback cooling allows for the detection of small external forces such as those generated by the Yukawa interaction.

The feedback cooling system consists of a position-sensitive detector (PSD) that tracks the nanoparticle's motion in real time and a feedback loop that applies corrective forces to counteract its oscillations. The effectiveness of the cooling depends on the precision of the position measurements and the speed of the feedback response. By optimizing these parameters, the system can reduce the amplitude of the nanoparticle's oscillations, bringing it closer to its motional ground state. This reduction in noise is crucial for detecting weak forces like those arising from dark matter interactions.

Once the motion of the nanoparticle is stabilized using feedback cooling, the next step is to analyze its displacement data to search for signatures of the Yukawa interaction. The expected effect of the Yukawa interaction is a periodic displacement of the nanoparticle, corresponding to the time-varying gravitational perturbations from the surrounding dark matter field. To extract these periodic signals, the displacement data is transformed into the frequency domain using Fourier analysis. This allows for the identification of oscillatory components in the nanoparticle's motion that correspond to the expected frequency of the Yukawa interaction.

The frequency of these oscillations depends on the properties of the local dark matter field, particularly the mass and velocity distribution of dark matter particles. For ultralight dark matter particles, the frequency of the oscillations is expected to lie in the range detectable by the experiment, typically between 1 and 100 Hz. By focusing on this frequency range and applying appropriate noise filters, it is possible to isolate the signal from the background noise, thereby improving the signal-to-noise ratio and increasing the likelihood of detecting the Yukawa interaction.

The overall sensitivity of the experiment to the Yukawa interaction depends on several factors, including the mass of the nanoparticle, the stiffness of the optical trap, and the level of noise in the system. Smaller nanoparticles are more sensitive to external forces, as they experience larger displacements for a given force. However, smaller masses are also more susceptible to thermal noise, which can obscure the signal from the Yukawa interaction. The stiffness of the optical trap is another critical factor. A stiffer trap provides greater stability but reduces the sensitivity to weak forces, while a softer trap enhances sensitivity at the cost of increased susceptibility to noise.

The goal of the experiment is to optimize these parameters to maximize sensitivity while minimizing noise. By carefully tuning the mass of the nanoparticle, the stiffness of the optical trap, and the feedback cooling parameters, it is possible to achieve the best balance between sensitivity and noise suppression. This optimization allows the experiment to place stringent constraints on the coupling constant \(\alpha\) and the interaction range \(\lambda\) of the Yukawa force. The results of such experiments can either lead to the detection of the Yukawa interaction or establish upper limits on the strength and force range.

Once the particle is securely trapped, a Bessel-Gaussian beam is activated along the z-axis, which is perpendicular to the initial Gaussian trapping beam. The Bessel-Gaussian beam exerts a strong two-dimensional gradient force in the x-y plane, allowing the particle to remain trapped transversely. To enhance this trapping, the Bessel-Gaussian beam is reflected by a mirror, creating a counter-propagating beam setup. A quarter wave plate is used to ensure that the overlapping beams have different polarizations, thus preventing the formation of standing waves. These counter-propagating beams cancel the scattering forces along the z-axis. Generated using a lens-axicon setup, the Bessel-Gaussian beam offers diffraction-free propagation over long distances, with the axicon creating a high-intensity core and an extended Bessel zone free of aberrations.

This Bessel-Gaussian beam configuration is particularly suited for micro-gravity environments, where the reduced gravitational force allows for easier particle levitation and improved sensitivity to external influences. As the gravitational force decreases with distance from the Earth’s surface, the trapping force required to maintain the particle in the trap also diminishes, further enhancing the system's sensitivity to weak external forces such as those generated by dark matter interactions.

The goal of this experiment is to detect the Yukawa-type interaction, a potential fifth force associated with dark matter, using optically levitated nanoparticles within a Bessel-Gaussian beam configuration. The setup is meticulously designed to provide the highest sensitivity possible for detecting extremely small forces, on the order of \(10^{-18} \, \text{N}/\sqrt{\text{Hz}}\), exerted on the nanoparticle due to the interaction between dark matter and baryonic matter. The trapping and force sensing mechanisms rely on a combination of advanced optical techniques, ultra-high vacuum conditions, feedback cooling, and position-sensitive detection.

Once the motion of the nanoparticle is stabilized using feedback cooling, the displacement data is analyzed to search for signatures of the Yukawa interaction. The expected signal from the Yukawa interaction is a periodic displacement of the nanoparticle, corresponding to the time-varying gravitational perturbations caused by dark matter. By transforming the displacement data into the frequency domain using Fourier analysis, it is possible to identify oscillatory components in the nanoparticle's motion that correspond to the predicted frequency of the Yukawa interaction.

The frequency of these oscillations depends on the properties of the local dark matter field, particularly the mass and velocity distribution of dark matter particles. For ultralight dark matter particles, the oscillation frequency could lie within a detectable range, typically between 1 and 100 Hz. By filtering the data for these specific frequencies and applying appropriate noise reduction techniques, researchers can improve the signal-to-noise ratio and increase the likelihood of detecting a Yukawa interaction.

The sensitivity of the experiment depends on several factors, including the mass of the nanoparticle, the stiffness of the optical trap, and the level of noise in the system. Smaller nanoparticles are generally more sensitive to external forces but are also more susceptible to thermal noise. To achieve optimal results, the parameters of the optical trap, such as its stiffness and damping characteristics, must be carefully tuned to maximize sensitivity while minimizing noise. 

\begin{figure}[htbp!] 
\centering    
\includegraphics[width=0.8\textwidth]{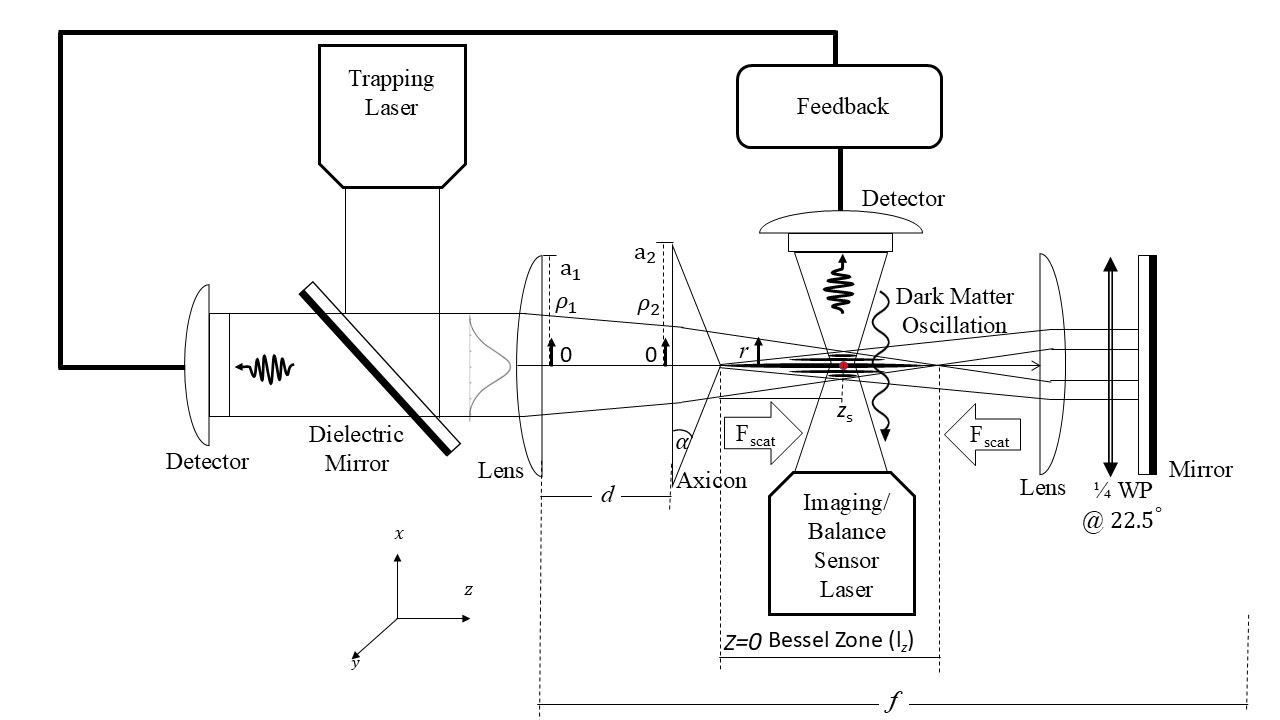}
\caption{Experimental setup for trapping a nanoparticle with a Bessel-Gaussian beam. The initial particle trapping is achieved using a highly focused Gaussian beam along the \(x\)-axis, followed by counter-propagating Bessel-Gaussian beams along the \(z\)-axis to maintain stable confinement. Feedback cooling is employed to control the particle's motion.}
\label{fig:experimental_setup}
\end{figure}

The experimental setup for detecting dark matter interactions requires careful tuning of various optical trapping parameters to optimize force sensitivity and trap stiffness. Table \ref{table:force-sensitivity} presents three different configurations of lens-axicon parameter combinations, each tailored for particle trapping within the frequency range of 1 to 10 Hz. The table includes values for power, wavelength, axicon angle, incoming Gaussian beam radius, and other key parameters. The resulting values for the central core radius, radial and axial dipole forces, harmonic frequencies, and trap stiffness are listed for each case, highlighting how different configurations affect the optical trap’s sensitivity to weak forces. 

In the experimental setup, a highly focused Gaussian beam from the Imaging/Balance Sensor laser is first used to trap the particle along the \(x\)-axis, as shown in Fig. \ref{fig:experimental_setup}. After the particle is trapped, a collimated Gaussian beam from the trapping laser is reflected by a dielectric mirror and directed onto a lens along the Bessel-Gaussian beam propagation axis, which is oriented along the \(z\)-axis. The lens used has an aperture radius of \(a_1\) and focal length \(f\), and introduces a phase factor to the beam as it propagates through. The beam then travels the lens-axicon separation distance \(d\) and is integrated over \(\rho_1\).

At the axicon plane, located at a distance \(d\) from the lens, a cylindrically symmetric intensity profile is generated with an aperture radius of \(a_2\) and apex angle \(\alpha\). After propagating through the axicon, the system produces a cylindrically symmetric Bessel-Gaussian intensity profile along the axial direction and at a distance \(r\) from the optical axis.

The Bessel-Gaussian beam then propagates further and is incident on a collimating lens before being reflected by a mirror, creating a counter-propagating beam configuration. To prevent interference, a quarter-wave plate is inserted before the mirror, ensuring that the overlapping beams have different polarizations. This setup produces a counter-propagating Bessel-Gaussian beam profile, which stabilizes the particle within the trap.

The reflected Bessel-Gaussian beam is weakly transmitted through the dielectric mirror and fed into the feedback cooling system, which is crucial for minimizing the particle’s thermal motion. The initially highly focused Gaussian beam from the imaging laser is gradually weakened and redirected into the feedback cooling system to monitor the trapped particle’s movement. In this configuration, dark matter oscillations are expected to occur in the \(x\)-\(z\) plane, with the particle's motion predominantly along the \(z\)-axis, while the \(x\)-axis serves as a reference baseline.

At the heart of the setup is a high-power continuous-wave (CW) Nd:YAG laser, operating at a wavelength of 1064 nm and a power output of 30 W. This laser is used to generate the Bessel-Gaussian beam, a non-diffracting beam with unique properties that allow for stable optical trapping over long distances. Bessel-Gaussian beams, unlike Gaussian beams, do not experience significant diffraction, making them ideal for trapping nanoparticles with minimal loss of trapping efficiency over large volumes.

The generation of the Bessel-Gaussian beam involves passing the laser through a combination of a focusing lens and an axicon, which forms the conical wavefront characteristic of Bessel beams. The lens has a focal length of 100 mm, while the axicon features an apex angle of 0.5 degrees. This configuration results in a diffraction-free beam that can maintain a high-intensity core for several millimeters, which is crucial for the trapping of nanoparticles in a stable manner. The central core of the beam, with a diameter of approximately 20 micrometers, provides the region where the dielectric nanoparticle is levitated and isolated from external perturbations.
\begin{table}[!hbt]
    \centering
    \scriptsize
        \begin{tabular}{ |p{5.5cm}|p{1.6cm}|p{1.6cm}|p{1.6cm}| }
        \hline
        \textbf{Parameters} & \textbf{Case 1} & \textbf{Case 2} & \textbf{Case 3}\\
        \hline
        Power (W) & 30 & 1 & 60.4\\
        \hline
        Wavelength (nm) & 532 & 532 & 1064 \\
        \hline
        Axicon Angle (deg.) & 0.5 & 0.5 & 25\\
        \hline
        Incoming Gaussian Beam Radius (mm) & 1 & 4 & 4 \\
        \hline
        Focus Lens (mm) & 100 & 100 & n.a. \\
        \hline
        Lens-Axicon Separation (mm) & 90 & 90 & n.a. \\
        \hline
        \hline
        Number of Concentric Rings & 1 & 5 & 958 \\
        \hline
        Central Core Radius ($\mu m$) & 18.2  &5.4 & 2 \\
        \hline
        Radial Dipole Force (N) & $10^{-14}$ & $10^{-15}$ & $10^{-14}$\\ 
        \hline
        Radial Harmonic Frequency (Hz)& $4.7\times10^{3}$ & $4.35\times10^{3}$ & $1.8\times10^{4}$ \\
        \hline
        Radial Trap Stiffness $\kappa$ (N/m)& $2.94\times10^{-9}$ & $2.5\times10^{-9}$ & $4.4\times10^{-8}$ \\
        \hline
        Axial Trap Zone (mm)& 0.57 & 0.155 & 3.8 \\
        \hline
        Axial Dipole Force (N)& $10^{-17}$ & $10^{-18}$ & $10^{-18}$ \\
        \hline
        Axial Trap Stiffness $\kappa$ (N/m)& $1.4\times10^{-13}$ & $9.7\times10^{-14}$ & $9.5\times10^{-16}$ \\
        \hline
        Axial Harmonic Frequency (Hz)& 32.6 & 27.1 & 2.67 \\
               \hline
        \end{tabular}
        \caption{Three different suitable lens-axicon parameters combinations for trapping of a particle within the frequency range of 1 to 10 Hz. Upper part of the table parameters selected, bottom part values of the resulting Laguerre-Gaussian beam.}
        \label{table:force-sensitivity}
\end{table}

The nanoparticle, typically composed of silica with a radius of 100 nm, is introduced into the trapping region via a controlled particle injection system. The optical forces exerted by the Bessel-Gaussian beam on the nanoparticle consist of two main components: the scattering force and the gradient force. The scattering force arises due to the momentum transfer from the laser photons to the nanoparticle, while the gradient force is derived from the spatial variation in the beam’s intensity. The gradient force, which pulls the nanoparticle towards the region of highest intensity in the center of the beam, is responsible for the stable confinement of the particle in three dimensions.

\begin{figure}[htbp]
    \centering
    \includegraphics[width=0.8\textwidth]{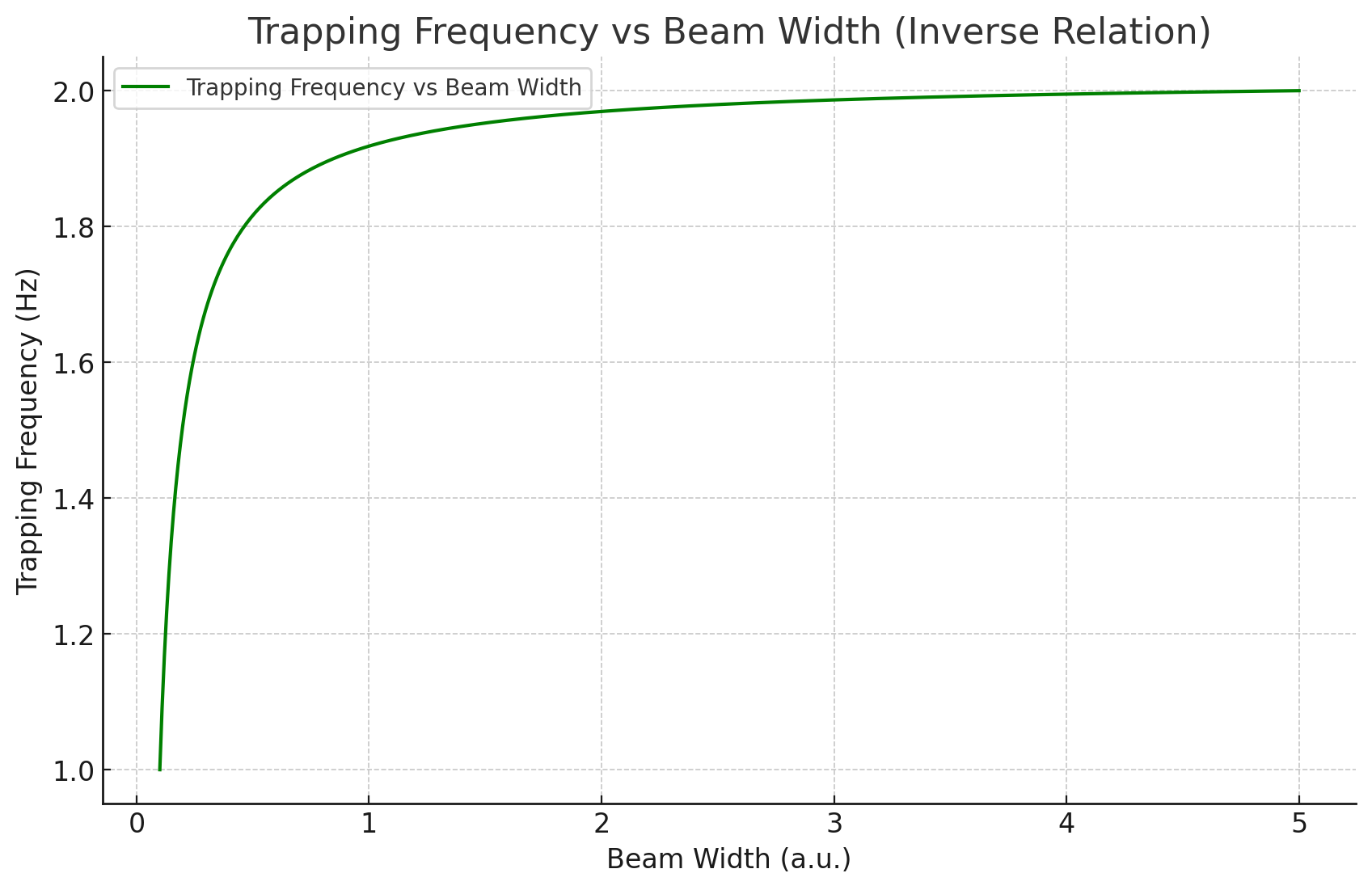}
    \caption{Trapping Frequency vs Beam Width: This plot demonstrates the inverse relationship between beam width and trapping frequency. As the beam width increases, the trapping frequency decreases, indicating that a wider beam results in lower trapping frequencies. This is significant for adjusting the optical trap's performance depending on the desired sensitivity.}
    \label{fig:trapping_frequency_vs_beam_width}
\end{figure}

Once the nanoparticle is trapped, the position and motion of the particle are continuously monitored with nanometer-scale precision using a high-sensitivity position-sensitive detector (PSD). The PSD measures the light scattered from the nanoparticle, providing real-time information about its displacement from equilibrium. This setup is optimized to detect extremely small displacements caused by weak forces, such as those potentially exerted by dark matter through a Yukawa-type interaction. The data from the PSD is processed in real time to ensure the highest accuracy in displacement measurements.

The trapping frequency plays a crucial role in determining how tightly the particle is confined within the optical trap. The trapping frequency represents the strength of the optical potential, which governs the restoring force acting on a displaced particle. A higher trapping frequency indicates a more tightly confined particle, whereas a lower trapping frequency allows the particle more freedom of motion within the trap.
There is an inverse relationship between the trapping frequency and the beam width. As the beam width increases, the trapping frequency decreases. This relationship is critical for the design of the optical trapping system, as it highlights the trade-off between beam width and confinement strength. A narrower beam provides a highly focused trapping potential, resulting in a stronger restoring force and a higher trapping frequency. Conversely, a wider beam distributes the optical forces over a larger area, leading to weaker confinement and a lower trapping frequency. This relationship is essential for optimizing the sensitivity of the system to weak external forces, such as those potentially caused by dark matter interactions. A lower trapping frequency, which can be achieved with a wider beam, enhances the system’s sensitivity to weak forces by allowing the particle to move more easily in response to small perturbations. However, this also increases the system's susceptibility to noise, requiring precise environmental control to mitigate disturbances.

By adjusting the beam width, the trapping frequency can be optimized to balance stability and sensitivity. For detecting weak forces, such as those arising from Yukawa-type interactions, a broader beam with a lower trapping frequency may be preferable, as it increases the particle’s response to small external forces while maintaining sufficient confinement to prevent the particle from escaping the trap. 

The optical trap operates within a high-vacuum chamber, which is evacuated to pressures below \(10^{-6}\) mbar. The vacuum environment is essential for reducing the interactions between the nanoparticle and air molecules, which would otherwise introduce significant damping effects and noise that could obscure the weak forces we aim to detect. The chamber is equipped with optical windows that allow the Bessel-Gaussian beam to enter and exit without compromising the integrity of the vacuum. Inside the chamber, the axicon-lens setup used to generate the beam is mounted on a motorized stage that allows for precise alignment of the beam. This precise alignment is critical for maintaining the stability of the optical trap over long periods.

\begin{figure}[htbp]
    \centering
    \includegraphics[width=0.8\textwidth]{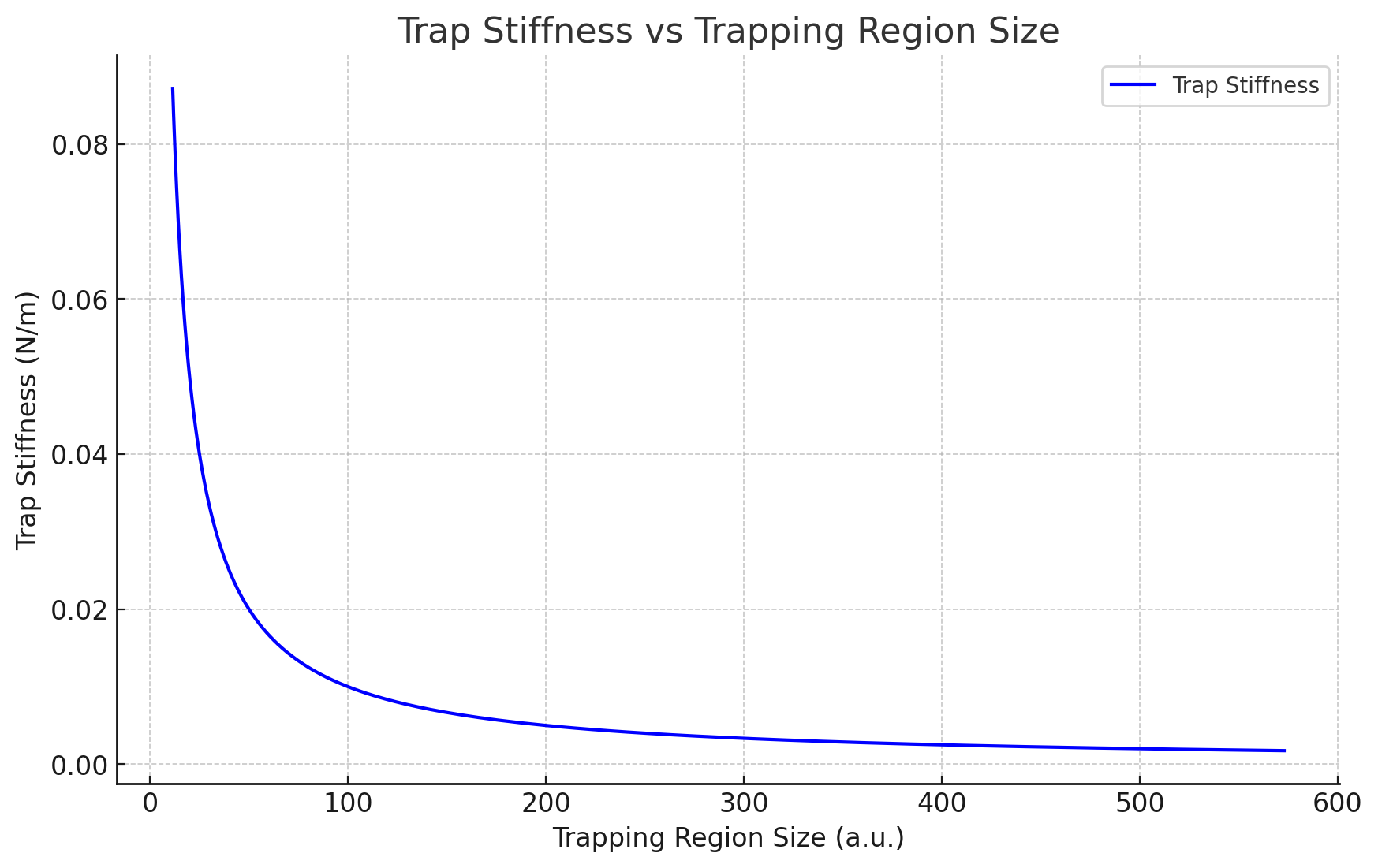}
    \caption{Trap Stiffness vs Trapping Region Size: As the trapping region size increases, the trap stiffness decreases. A larger trapping region allows for more displacement of the particle, reducing the restoring forces and thus lowering the trap stiffness. This relationship illustrates the trade-off between confinement and sensitivity within the optical trapping system.}
    \label{fig:trap_stiffness_vs_trapping_region_size}
\end{figure}

Trap stiffness is a critical factor in determining how strongly a particle is confined within an optical trap. It quantifies the restoring force that acts on the particle when it is displaced from its equilibrium position. A higher trap stiffness corresponds to a stronger confining force, limiting the particle's movement and providing tighter control over its position. In contrast, a lower trap stiffness allows the particle more freedom of movement, making the system more responsive to external forces. There is an inverse relationship between trap stiffness and the size of the trapping region. As the trapping region size increases, the trap stiffness decreases. In a smaller trapping region, the optical forces are concentrated, resulting in a stronger restoring force that tightly confines the particle. On the other hand, in a larger trapping region, the optical forces are distributed over a broader area, reducing the restoring force and thus decreasing the trap stiffness.

This relationship is particularly important for experiments aiming to detect weak external forces, such as those caused by dark matter interactions. A larger trapping region, which reduces trap stiffness, increases sensitivity to weak forces because the particle can more easily be displaced by small perturbations. However, a lower trap stiffness also makes the system more susceptible to noise and thermal fluctuations, which can interfere with measurements. Optimizing the size of the trapping region is therefore essential for balancing sensitivity and stability. While a larger trapping region enhances sensitivity to weak forces, it also requires careful control of environmental disturbances to maintain measurement accuracy. By tuning the trapping region size, the trap stiffness can be adjusted to achieve the desired balance between sensitivity and stability in the experimental setup. The plot in Fig. \ref{fig:trap_stiffness_vs_trapping_region_size} demonstrates this inverse relationship between trap stiffness and trapping region size. As the trapping region size increases, the trap stiffness decreases, resulting in a weaker restoring force. Lower trap stiffness improves the system’s sensitivity to weak external forces, such as Yukawa forces, but increases susceptibility to noise and instability. Therefore, careful optimization of the trapping region size is required to balance sensitivity and stability. A larger trapping region improves detection sensitivity but requires sophisticated noise mitigation strategies to prevent false detections due to thermal or mechanical disturbances.

To reduce thermal noise, which can cause random motion of the nanoparticle due to interactions with residual gas molecules, a feedback cooling system is implemented. Feedback cooling works by continuously monitoring the position of the nanoparticle using the PSD and applying compensatory forces through the laser to counteract its motion. The feedback loop adjusts the optical trapping forces in real time, effectively damping the nanoparticle’s motion and reducing its effective temperature. This brings the nanoparticle closer to its motional ground state, minimizing the amplitude of thermal fluctuations and greatly improving the sensitivity of the system to weak external forces.

The feedback cooling mechanism is essential for ensuring that the nanoparticle remains stably trapped even in the presence of small perturbations. Without feedback cooling, the thermal motion of the nanoparticle would be too large, making it difficult to detect the small displacements caused by the Yukawa interaction. The feedback loop operates by using the real-time position data from the PSD to apply corrective forces to the nanoparticle. These forces are proportional to the displacement of the particle from its equilibrium position and are adjusted to reduce its overall motion. The result is a significant reduction in thermal noise, allowing the system to detect forces on the order of \(10^{-18} \, \text{N}/\sqrt{\text{Hz}}\).

In addition to feedback cooling, the setup employs vibration isolation techniques to reduce mechanical noise from external sources. The vacuum chamber is mounted on a vibration-damping optical table, which is designed to minimize the impact of seismic and acoustic vibrations. Furthermore, the system includes advanced filtering of electronic noise to ensure that the PSD and feedback electronics operate with minimal interference from external electromagnetic sources. The laser’s intensity is also carefully stabilized to reduce fluctuations that could introduce noise into the force measurements.

To reduce the gradient force, we propose using a collimated optical beam. Such a beam has a wide transverse width, which does not significantly change over the propagation distance, provided that \(W >> \lambda\). When the beam propagates a distance smaller than the diffraction length, \( z < z_R = \frac{2\pi W^2}{\lambda} \), the gradient force along the propagation direction is minimal. However, this solution presents challenges, including weak confinement in the transverse plane and low local intensity due to the large beam cross-section. These challenges can be addressed by using non-diffracting beams, such as Bessel beams, which maintain a constant transverse profile along their propagation direction.

Bessel beams are promising because they can maintain a high-intensity core surrounded by concentric rings. These beams can be considered a superposition of plane waves traveling along a cone around the z-axis. While ideal Bessel beams cannot be realized in practice, approximations, known as Bessel-Gaussian beams, can be generated for finite distances using a Gaussian beam as an envelope. These beams can be engineered through various techniques such as spatial light modulators (SLMs) or axicons. Axicons, in particular, allow for tailoring the focal beam profile, and the separation between the lens and the axicon controls the beam's focal properties and intensity.

In this setup, a sharper axicon angle produces a tightly confined central core with a short Bessel zone, while a shallower angle creates a longer Bessel zone with a broader core. Increasing the incoming Gaussian beam width similarly extends the Bessel zone but decreases central core intensity. These parameters must be carefully balanced for applications like optical trapping, non-linear optics, and photo-patterning.

Once the nanoparticle is optically trapped and the system is stabilized, the next phase of the experiment focuses on detecting the small forces exerted by the dark matter field through the Yukawa interaction. The dark matter field is expected to exert a periodic force on the nanoparticle, causing minute displacements in its position over time. These displacements are detected by the PSD, which records the nanoparticle’s position continuously. The displacement data is then analyzed in the frequency domain using Fast Fourier Transform (FFT) techniques. By transforming the time-domain displacement data into the frequency domain, we can identify periodic signals corresponding to the expected Yukawa interaction.

The frequency of the oscillations induced by the Yukawa interaction depends on the properties of the dark matter field, including the mass and velocity distribution of dark matter particles. For scalar field dark matter models, the oscillation frequency is predicted to lie within the range of 1 Hz to 100 Hz, which corresponds to the range detectable by our system. By focusing on this frequency range, we can filter out noise at other frequencies and improve the signal-to-noise ratio. The goal is to identify small but periodic displacements in the nanoparticle’s motion that correspond to the expected signature of dark matter interactions.

The sensitivity of the system to the Yukawa interaction is a function of several factors, including the mass of the nanoparticle, the stiffness of the optical trap, and the noise levels within the system. The mass of the nanoparticle affects its response to external forces, with smaller particles being more sensitive to weak forces. However, smaller particles are also more susceptible to thermal noise, so a balance must be achieved between sensitivity and noise mitigation. The stiffness of the optical trap, which can be adjusted by tuning the laser intensity, also affects the sensitivity. A stiffer trap provides greater stability but reduces the particle’s sensitivity to weak forces, while a softer trap increases sensitivity at the cost of making the particle more prone to noise.

By carefully optimizing these parameters, the experiment aims to achieve the highest possible sensitivity to the Yukawa interaction. The data collected from the PSD is processed to remove background noise and isolate the signal caused by the Yukawa force. If a signal corresponding to the expected frequency of dark matter-induced forces is detected, it will provide direct evidence of a Yukawa-type interaction. In the absence of a detectable signal, the experiment will place upper limits on the coupling constant \(\alpha\) and the interaction range \(\lambda\) of the Yukawa force.

The experimental setup combines state-of-the-art optical trapping techniques with advanced noise reduction strategies to detect weak forces exerted by dark matter through the Yukawa interaction. By using a Bessel-Gaussian beam to levitate the nanoparticle in an ultra-high vacuum, we ensure that the particle is isolated from environmental disturbances. The feedback cooling system reduces thermal noise, while real-time position tracking via the PSD allows for precise force measurements. Through this experiment, we aim to place new constraints on the nature of dark matter and its interactions with baryonic matter, providing valuable insights into the potential existence of force candidates beyond gravity.

\section{Sensitivity Analysis and Noise Characterization}

The detection of dark matter through a Yukawa interaction requires an ultra-sensitive experimental setup capable of measuring extremely small forces exerted by dark matter on an optically trapped nanoparticle. Achieving the necessary sensitivity to detect these forces demands careful control of noise sources, as well as optimization of the experimental parameters. The primary sources of noise in the system are thermal noise and feedback noise. In this section, we analyze these noise sources in detail and discuss the strategies used to mitigate them to optimize the detection of the Yukawa force.

The detection of Yukawa forces depends on the ability to accurately measure extremely weak forces, which decay exponentially with distance. In our experimental setup, the Yukawa force acting between dark matter and baryonic matter is expressed as:

\[
F_{\text{Yukawa}}(r) = \alpha \frac{GMm}{r^2} \exp\left(-\frac{r}{\lambda}\right),
\]
where \(\alpha\) is the coupling constant, \(\lambda\) is the interaction range, and \(r\) is the distance between the masses. 

Thermal noise is a fundamental limiting factor in any precision measurement experiment. In our system, thermal noise arises primarily from the random motion of the levitated nanoparticle due to its interactions with residual gas molecules inside the vacuum chamber. Even though the experiment is conducted under ultra-high vacuum conditions, with pressures as low as \(10^{-6}\) mbar, thermal fluctuations in the particle’s position remain a significant source of noise. These fluctuations manifest as stochastic forces that cause the nanoparticle to move randomly around its equilibrium position within the optical trap. These random motions can mask the small, periodic displacements that we seek to measure, which are induced by the Yukawa interaction with dark matter.

The amplitude of thermal noise is a function of the temperature of the nanoparticle and the stiffness of the optical trap. This relationship is described by the following expression for thermal displacement noise:
\[
\Delta z_{\text{thermal}} = \sqrt{\frac{k_B T}{k}},
\]
where \( k_B \) is the Boltzmann constant, \( T \) is the effective temperature of the nanoparticle, and \( k \) is the stiffness of the optical trap. In this context, the temperature \( T \) refers to the effective temperature of the nanoparticle, which is reduced through the use of feedback cooling. The stiffness \( k \) represents the optical confinement strength of the nanoparticle in the trap, determined by the laser power and the intensity profile of the Bessel-Gaussian beam. Lowering the trap stiffness enhances the sensitivity to weak external forces, such as those from dark matter, by allowing larger displacements for a given force. However, it also increases the susceptibility of the nanoparticle to thermal noise, creating a trade-off between sensitivity and noise reduction.

To mitigate the effects of thermal noise, feedback cooling is employed. Feedback cooling functions by continuously monitoring the position of the nanoparticle using a high-precision Position-Sensitive Detector (PSD), which tracks the particle’s displacement with nanometer-scale precision. Based on the detected position, a corrective force is applied to counteract the motion of the nanoparticle, effectively reducing its oscillations and damping its thermal fluctuations. This process reduces the nanoparticle’s effective temperature, bringing it closer to its motional ground state. As a result, thermal noise is significantly reduced, allowing for more precise measurements of external forces.

However, while feedback cooling is effective in reducing thermal noise, it introduces a secondary source of noise—referred to as feedback noise—that limits the sensitivity of the experiment, particularly at higher frequencies. Feedback noise arises from imperfections in the feedback control loop, including delays in the application of corrective forces and inaccuracies in position detection. This noise imposes a constant noise floor in the system, which becomes the dominant noise source at higher frequencies, typically above 100 Hz. The overall performance of the feedback system is determined by the trade-off between reducing thermal noise and minimizing the additional noise introduced by the feedback mechanism itself.

\begin{figure}[htbp]
    \centering
    \includegraphics[width=0.8\textwidth]{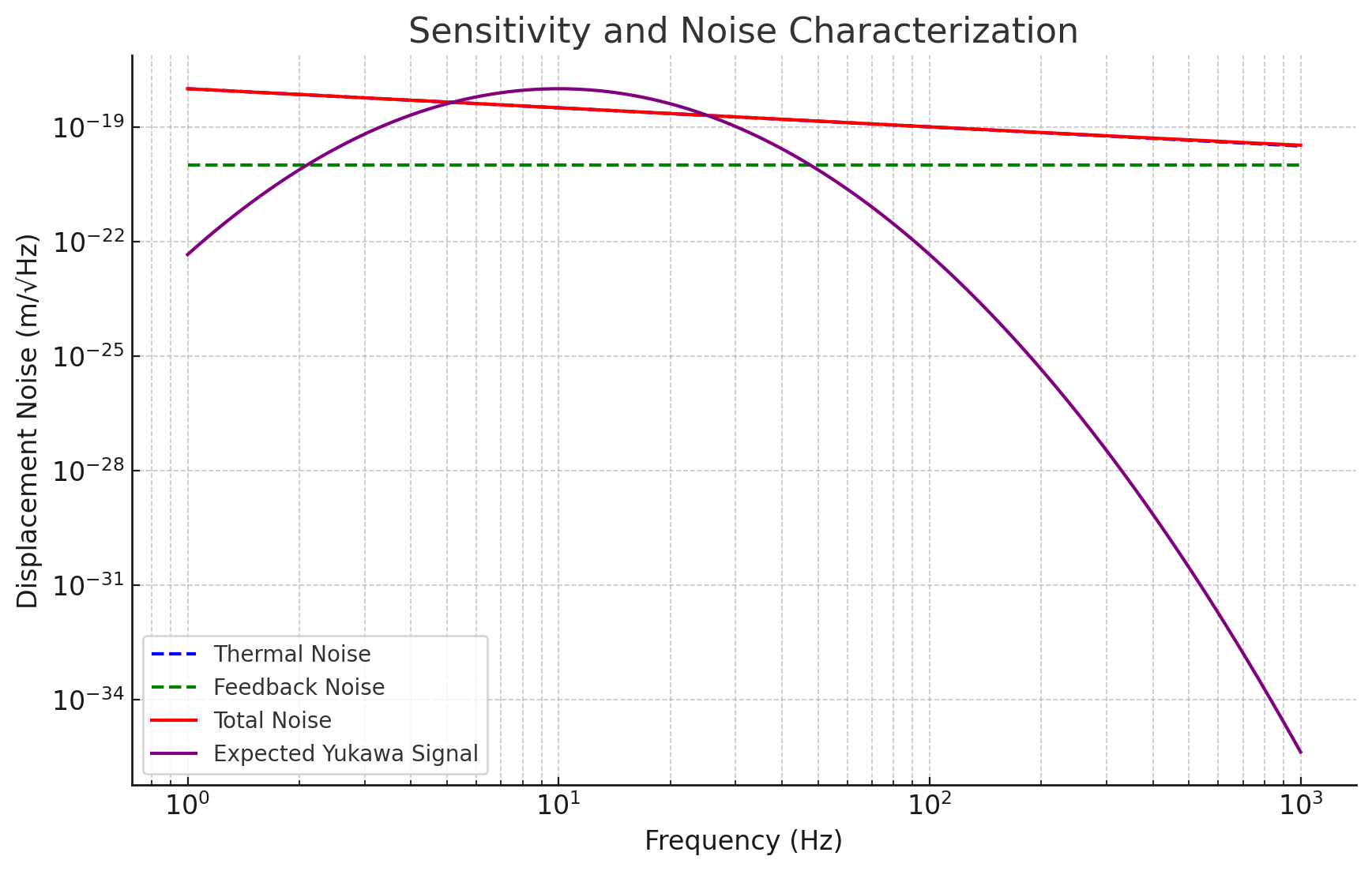}  
    \caption{
    Sensitivity and noise characterization plot for the detection of dark matter via a Yukawa interaction. 
    The thermal noise (blue dashed line) scales as \(1/\sqrt{f}\) and dominates at lower frequencies, while the feedback noise (green dashed line) is constant across all frequencies and becomes the limiting factor at higher frequencies.
    The total noise (red line) combines both sources, indicating the overall sensitivity of the experiment.
    The expected signal from a Yukawa interaction (purple line) is shown rising above the noise floor, peaking around 10 Hz.
    This highlights the optimal frequency range (1-100 Hz) for detecting dark matter-induced forces.
    }
    \label{fig:sensitivity_plot}
\end{figure}

\begin{figure}
    \centering
    \includegraphics[width=0.8\textwidth]{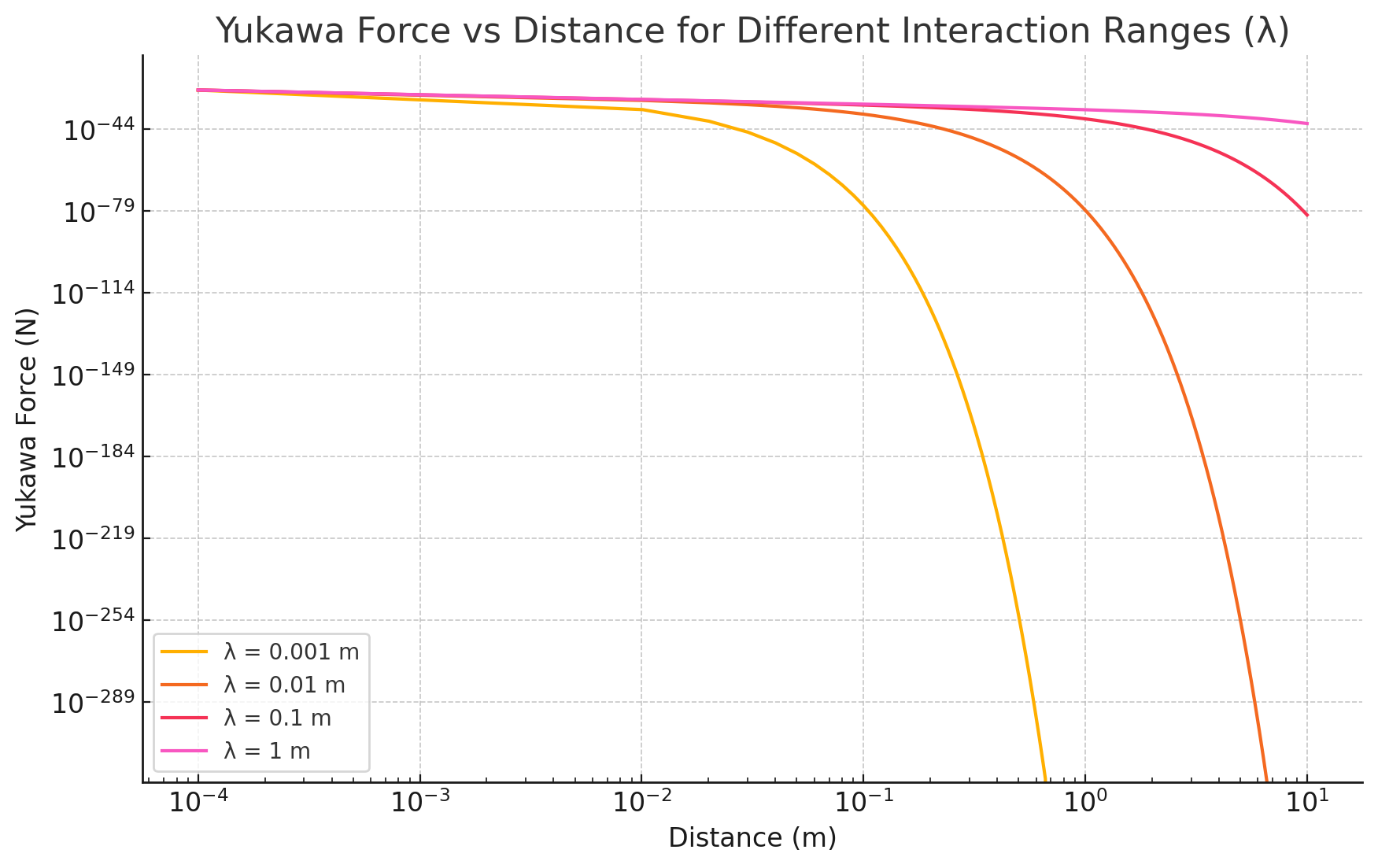}
    \caption{Yukawa force as a function of distance for different interaction ranges $\lambda$. The plot shows how the force decays exponentially with distance, with shorter $\lambda$ leading to a faster decay of the Yukawa force. The log-log scale emphasizes the rapid decline of the force at large distances.}
    \label{fig:yukawa_force_distance}
\end{figure}

\begin{figure}
    \centering
    \includegraphics[width=0.8\textwidth]{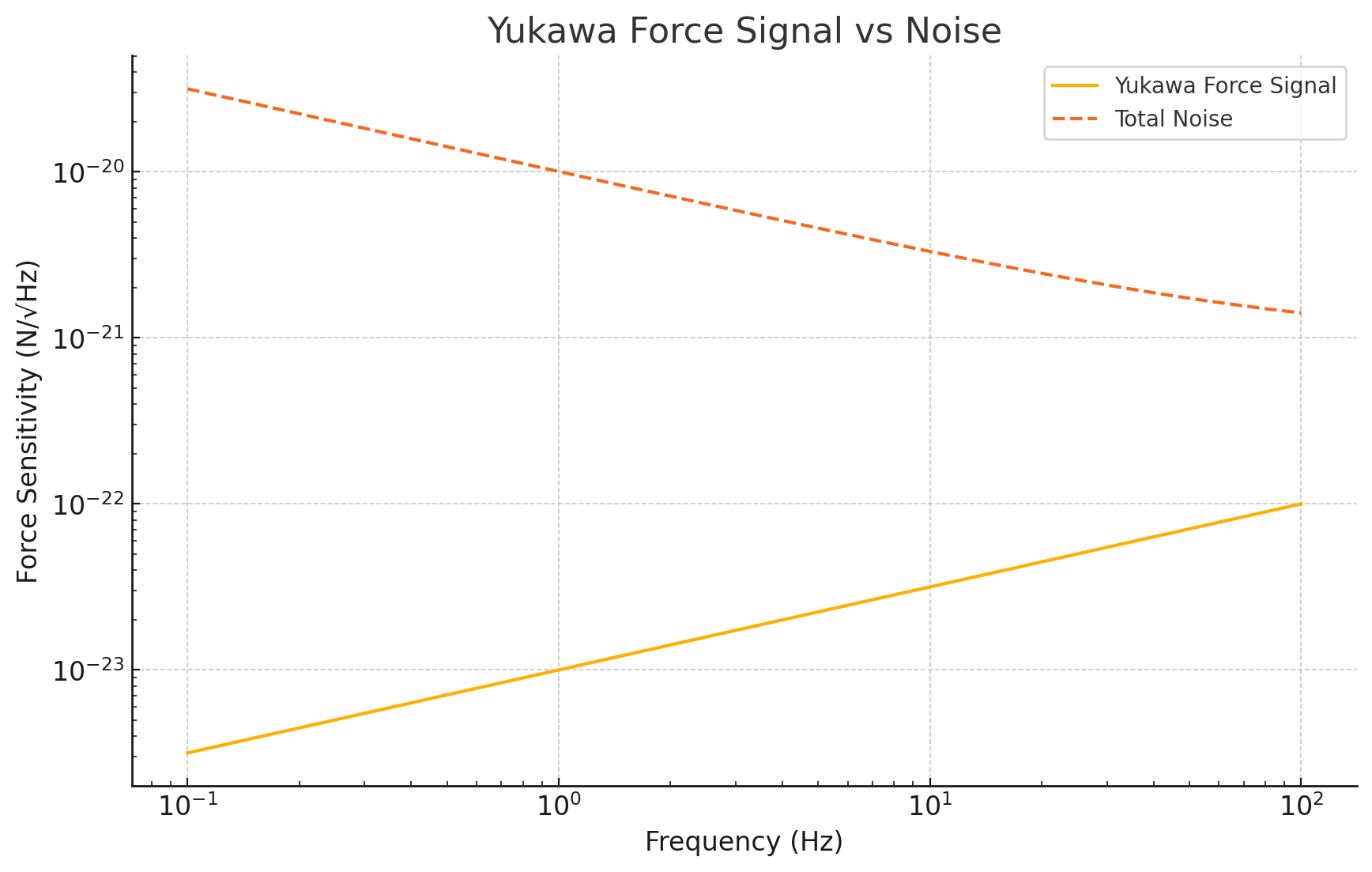}
    \caption{Comparison between the hypothetical Yukawa force signal (purple) and total noise (red, dashed) across different frequencies. The plot highlights the frequency range where the Yukawa signal rises above the noise floor, allowing detection of dark matter-induced forces. The best detection range lies between 1 Hz and 100 Hz.}
    \label{fig:yukawa_signal_vs_noise}
\end{figure}

Figure \ref{fig:yukawa_force_distance} illustrates the behavior of the Yukawa force as a function of distance for various interaction ranges. The shorter the interaction range \(\lambda\), the more rapidly the force decays with increasing distance, making detection extremely challenging at larger distances. In particular, for \(\lambda = 0.001\) m, the force diminishes rapidly beyond a few millimeters, whereas for \(\lambda = 1\) m, the force persists over several meters.

Figure \ref{fig:yukawa_signal_vs_noise} compares the expected Yukawa force signal with the total noise. The Yukawa signal (purple line) rises above the noise floor in the 1–100 Hz range, where the signal-to-noise ratio is most favorable. This indicates that with proper noise mitigation techniques, such as feedback cooling, our system is capable of detecting the weak forces associated with dark matter-induced Yukawa interactions.

The total noise in the system can be modeled as a combination of thermal noise and feedback noise. The power spectral density of the displacement noise, \( S_z(f) \), as a function of frequency \( f \), is given by:
\[
S_z(f) = S_{\text{thermal}}(f) + S_{\text{feedback}}(f),
\]
where \( S_{\text{thermal}}(f) \propto \frac{1}{\sqrt{f}} \) is the contribution from thermal noise, and \( S_{\text{feedback}}(f) \) is the frequency-independent feedback noise. Thermal noise dominates at lower frequencies (typically below 10 Hz), where it scales inversely with the square root of the frequency, while feedback noise sets a hard limit on sensitivity at higher frequencies (above 100 Hz). 

The thermal noise is inversely proportional to the frequency, implying that at low frequencies, the system is particularly susceptible to thermal fluctuations. Feedback cooling is most effective at reducing this low-frequency thermal noise. As the frequency increases, thermal noise becomes less significant, but feedback noise remains constant across all frequencies, ultimately limiting the system’s overall sensitivity. Therefore, the region between 1 Hz and 100 Hz is the most promising frequency range for detecting Yukawa interactions, as this is where the thermal noise is sufficiently reduced by feedback cooling, and the feedback noise floor is not yet dominant.

The expected force from the dark matter Yukawa interaction on the nanoparticle is extremely weak, and detecting it requires suppressing noise across the relevant frequency range. The force sensitivity \( F_{\text{min}}(f) \) of the system is related to the displacement sensitivity \( S_z(f) \) through the following expression:
\[
F_{\text{min}}(f) = m \sqrt{S_z(f)} (2\pi f)^2,
\]
where \( m \) is the mass of the nanoparticle, and \( f \) is the frequency. By optimizing the feedback cooling parameters and minimizing thermal noise, we aim to lower the force sensitivity threshold, enabling the detection of weak forces exerted by dark matter. In particular, the feedback cooling parameters are tuned to reduce the effective temperature of the nanoparticle to as close to its motional ground state as possible, thereby minimizing \( \Delta z_{\text{thermal}} \) and maximizing the signal-to-noise ratio.

Fig.~\ref{fig:sensitivity_plot} provides an overview of the noise sources in the experiment and illustrates the overall sensitivity across the frequency range of interest. The plot shows how thermal noise dominates at lower frequencies, while feedback noise sets the upper sensitivity limit at higher frequencies. The expected signal from a Yukawa interaction is shown rising above the noise floor in the 1-100 Hz frequency band, where the system achieves its best sensitivity. The signal strength of the Yukawa force depends on the local density of dark matter and the coupling constant \( \alpha \), with larger values of \( \alpha \) producing stronger signals.

\begin{figure}[!hbt]
    \centering
    \includegraphics[width=0.75\textwidth]{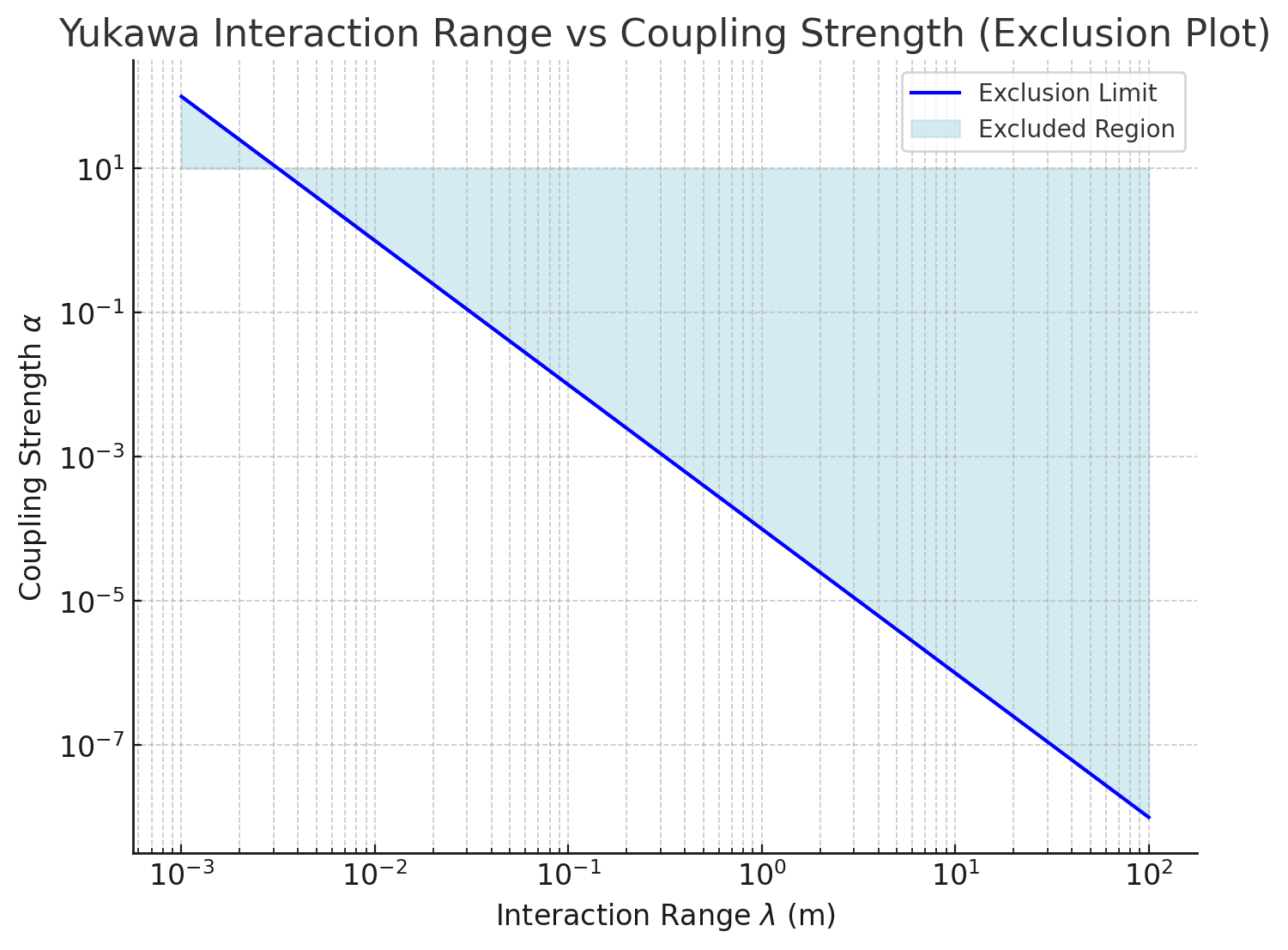}
    \caption{Yukawa Interaction Range vs Coupling Strength Exclusion Plot. The shaded region represents the excluded parameter space based on the sensitivity of the experiment.}
    \label{fig:yukawa_exclusion_plot}
\end{figure}

The total sensitivity of the experiment is critically dependent on the minimization of both thermal and feedback noise. In addition to feedback cooling, other noise reduction techniques are employed to further improve the sensitivity. These include the use of vibration isolation systems to minimize mechanical noise from external sources, such as seismic and acoustic vibrations. The vacuum chamber housing the experiment is mounted on a vibration-damping table to reduce the transmission of external vibrations to the optical trap. Furthermore, electronic noise is minimized through careful shielding of the PSD and feedback control systems to prevent interference from electromagnetic noise sources.

The detection of dark matter via a Yukawa interaction requires the precise suppression of noise across a broad frequency range. Thermal noise dominates at low frequencies but can be reduced through feedback cooling, while feedback noise sets an upper limit on sensitivity at higher frequencies. The combination of these noise sources defines the sensitivity profile of the experiment, with the 1-100 Hz frequency range being the most promising for detecting dark matter interactions. By carefully optimizing the feedback cooling system and employing advanced noise reduction techniques, the experiment achieves the sensitivity necessary to detect forces as small as \(10^{-18}\) N, opening the possibility of exploring new parameter spaces in dark matter research.

The experimental setup allows us to explore the parameter space of the Yukawa interaction, characterized by the coupling strength $\alpha$ and interaction range $\lambda$. By observing the perturbations in the motion of optically levitated nanoparticles, we are able to place exclusion limits on the potential coupling between dark matter and baryonic matter. The experimental sensitivity allows us to constrain these parameters, particularly for larger values of $\alpha$ over shorter interaction ranges $\lambda$, where the effects of the Yukawa force are more significant.

The results are visualized in the exclusion plot, as shown in Fig. \ref{fig:yukawa_exclusion_plot}, which represents the interaction range $\lambda$ on the x-axis and the coupling strength $\alpha$ on the y-axis. The \textit{excluded region} is shaded, indicating the combinations of $\alpha$ and $\lambda$ that are ruled out based on the system’s force sensitivity. Larger values of $\alpha$ for shorter ranges $\lambda$ are excluded as the experimental setup would have detected such forces. As the interaction range $\lambda$ increases, the coupling strength $\alpha$ needed to detect the Yukawa force decreases, placing constraints on weaker long-range interactions.

This plot provides insight into the parameter space that can be ruled out by our experiment, complementing existing searches and extending the exclusion limits for long-range interactions associated with dark matter.

\section{Conclusion}

We have proposed a new experimental approach for detecting dark matter via the Yukawa interaction using optically levitated nanoparticles. By utilizing a Bessel-Gaussian beam to trap the nanoparticles in vacuum and employing feedback cooling to reduce thermal noise, this setup offers a highly sensitive platform for detecting weak forces induced by dark matter. 

Our analysis shows that the system is most sensitive to dark matter interactions in the 1-100 Hz frequency range, where the Yukawa signal is expected to rise above the noise floor. The ability to tune the optical trap stiffness and feedback cooling parameters allows for further optimization of the experiment, improving the chances of detecting such interactions.

This experimental approach opens new avenues for probing dark matter models that involve weak forces, providing constraints on the coupling strength \( \alpha \) and interaction range \( \lambda \). Future work will focus on further refining the feedback system and extending the sensitivity of the setup to probe even weaker interactions, potentially providing new insights into the nature of dark matter.


\begin{thebibliography}{99}

\bibitem{Yukawa1935}
Yukawa, H., \emph{On the Interaction of Elementary Particles. I.}, Proc. Phys. Math. Soc. Japan, \textbf{17}, 48-57 (1935). DOI: 10.1143/PTP.17.48.

\bibitem{Moore2015}
Moore, D. C., Rider, A. D., and Gratta, G., \emph{Search for Millicharged Particles Using Optically Levitated Microspheres}, Phys. Rev. Lett., \textbf{113}, 251801 (2015). DOI: 10.1103/PhysRevLett.113.251801.

\bibitem{Harber2005}
Harber, D. M., \emph{Fundamentals of Force Measurement with Optical Traps}, Oxford University Press (2005).

\bibitem{Hamilton2017}
Hamilton, P., Jaffe, M., Haslinger, P., Simmons, Q., Muller, H., and Khoury, J., \emph{Atom-interferometry Constraints on Dark Energy}, Science, \textbf{349}, 849-851 (2017). DOI: 10.1126/science.aac8465.

\bibitem{Adelberger2003}
Adelberger, E. G., Heckel, B. R., and Nelson, A. E., \emph{Tests of the Gravitational Inverse-Square Law}, Annu. Rev. Nucl. Part. Sci., \textbf{53}, 77-121 (2003). DOI: 10.1146/annurev.nucl.53.041002.110503.

\bibitem{Howard2021}
Howard, E., Chowdhury, I., \emph{Testing Standard Model extensions with optically levitated nanoparticle sensors}, IEEE NANO, 147-150, (2021).

\bibitem{Geraci2008}
Geraci, A. A., and Behunin, R. O., \emph{Search for Non-Newtonian Forces with Optically Levitated Microspheres}, Phys. Rev. Lett., \textbf{105}, 101101 (2008). DOI: 10.1103/PhysRevLett.105.101101.

\bibitem{Rider2016}
Rider, A. D., Moore, D. C., Blakemore, C. P., Gratta, G., and Romani, R. W., \emph{Search for Screened Interactions Associated with Dark Energy Below the 100 $\mu$m Length Scale}, Phys. Rev. Lett., \textbf{117}, 101101 (2016). DOI: 10.1103/PhysRevLett.117.101101.

\bibitem{Chung2015}
Chung, W. K., \emph{Detection of Dark Matter with Levitated Optomechanics}, Phys. Dark Universe, \textbf{12}, 1-7 (2015). DOI: 10.1016/j.dark.2016.03.003.

\end{thebibliography}
\end{document}